\documentclass[runningheads,a4paper]{llncs}

\usepackage{amssymb}
\usepackage{graphicx}

\usepackage{url}
\urldef{\mailsa}\path|tomek@bci-lab.info|   
\newcommand{\keywords}[1]{\par\addvspace\baselineskip\noindent\keywordname\enspace\ignorespaces#1}

\begin{document}

\mainmatter  

\title{Two--step Input Spatial Auditory BCI for Japanese Kana Characters}

\titlerunning{Two--step Input Spatial Auditory BCI for Japanese Kana Characters}

%
%
\author{Moonjeong Chang\and Tomasz M. Rutkowski\thanks{The corresponding author.}}
\authorrunning{Moonjeong Chang and Tomasz M. Rutkowski}

\institute{Life Science Center of TARA, University of Tsukuba\\
1-1-1 Tennodai, Tsukuba, Ibaraki, Japan\\
\mailsa\\
\url{http://bci-lab.info/}}

%
%

\toctitle{Two--step Input Spatial Auditory BCI for Japanese Kana Characters}
\tocauthor{Moonjeong Chang and Tomasz M. Rutkowski}
\maketitle

\begin{abstract}
We present an auditory stimulus optimization and a pilot study of a two--step input speller application combined with a spatial auditory brain--computer interface (saBCI) for paralyzed users. The application has been developed for $45$, out of $48$ defining the full set, Japanese \emph{kana} characters in a two--step input procedure setting for an easy--to--use BCI--speller interface. The user first selects the representative letter of a subset, defining the second step. In the second step, the final choice is made. At each interfacing step, the choices are classified based on the P300 event related potential (ERP) responses captured in the EEG, as in the classic oddball paradigm. The BCI online experiment and EEG responses classification results of the pilot study confirm the effectiveness of the proposed spelling method.
\end{abstract}

\section{Introduction}

A brain-computer interface (BCI) is a technology that uses brainwaves only to control a computer, or a machine, without any body muscle movements~\cite{bciBOOKwolpaw}. This technology could allow disabled people, such as those suffering from locked--in syndrome (LIS), to regain communication skills or to manage daily life support related functions. In our research, we focus on an auditory and non--invasive BCI modality, since it allows for an application without any training even for the naive user. 

Among the state--of--the--art BCIs, the visual modality is the most successful because the evoked P300 responses usually have the largest amplitudes allowing for the easiest classification~\cite{MoonJeongBCImeeting2013}. However, it is known that some LIS users (e.g. advanced stage amyotrophic lateral sclerosis patients) cannot rely on visual modalities because they gradually lose intentional muscle control, including eye movements, focusing or intentional blinking. Our research hypothesis is that the spatial auditory BCI (saBCI)~\cite{iwpash2009tomek,moonjeongIICST2014} modality shall be a more suitable solution for establishing communication with LIS users and for creating the Japanese speller requiring at minimum the $45$ characters out for $48$ available for a smooth communication.
In order to do so, we propose to utilize the two--step input procedure to create an easy--to--use speller interface with a limited number of choices (commands) in each step. We also test various auditory stimulus scenarios using different combinations of female and male voices in the two--step input protocol.
We report results from a series of experiments in which the users were asked to spell short Japanese words using the two--step input scenarios.

The remainder of the paper is organized as follows. In the next section, we describe the online BCI experiment set--up, EEG acquisition, filtering and classification procedures, as well as the two--step input Japanese \emph{kana} characters' speller paradigm. Then the results of the online BCI experiments are discussed together with the P300 response latencies. Finally, conclusions are drawn and future research directions outlined.

\section{Methods}

\begin{figure}[!Hb]
	\centering
	\vspace{-0.25cm}
    \includegraphics[width=0.8\linewidth]{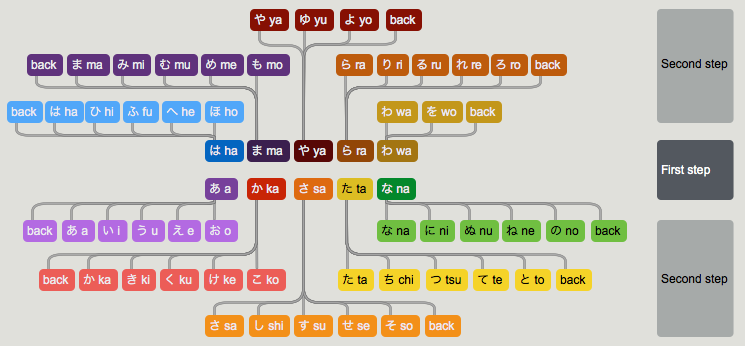}
    \caption{A diagram of the proposed two step Japanese \emph{kana} characters input procedure.}\label{fig.KANA}
\end{figure}

In the experiments reported in this paper, the Japanese \emph{kana} character set--based BCI--speller paradigm was investigated. The paradigm was designed to extend our previously reported pilot study~\cite{moonjeongIICST2014}.
In order to improve the previous $25$ characters--based paradigm, we designed the new interface as follows: the sound stimuli of male and the female voices were introduced in various input steps; ``a step--back'' command was added in the second step input using a tone of $440$~Hz in order to let the user to retreat from a possible error in the first step; in the first step the number of choices was increased from five to ten (see Figure~\ref{fig.KANA}). In order to create the spatial sound images with stimuli coming from distinct locations known to the user, we tested the following combinations. We used the sound stimulus patterns as follows: only the female voice modality with the same vocalization presented in the both input steps; male~$\rightarrow$~female voices modality in the first and second steps, respectively; female~$\rightarrow$~male voices modality with an opposite set--up to the previous one.
The speech occurrences were adopted from a sound dataset constructed of male and female natural voice recordings available for research purposed from~\cite{soundDatebase}. We used the sound stimuli of $45$~Japanese \emph{kana} characters.
The auditory stimuli were delivered from the ten distinct spatial locations using loudspeakers positioned in two horizontal lines in front of the user's head. The five loudspeakers were at the user ears' level and the remaining five in a line one meter above.
In order to manage the spatial sounds distributions and precise presentations to evoke the time aligned P300 responses an in--house multimedia application was developed in \emph{MAX} environment~\cite{maxMSP}. 
The MAX application communicated with the in--house extended EEG acquisition, filtering and classification environment BCI2000~\cite{bci2000book} using an UDP network protocol.
We conducted the online EEG experiments in order to verify a possibility to input the    set of Japanese \emph{kana} characters of the proposed spatial auditory BCI (saBCI) speller paradigm. 
We tested whether there were significant differences in accuracies among the proposed spatial sound stimulus patterns.
In the experiments, ten healthy users took part (mean age of $24.0$ years old, standard deviation of $0.63$ years). 
The online (real--time) EEG experiments were conducted in accordance with the \emph{WMA Declaration of Helsinki - Ethical Principles for Medical Research Involving Human Subjects}. 
The procedures were approved and designed in agreement with the ethical committee guidelines of the Faculty of Engineering, Information and Systems at University of Tsukuba, Japan.
Each user first conducted a short psychophysical test with a button press behavioral response to confirm understanding of the experimental set--up. Next, the EEG online BCI protocol experiments were conducted. In the EEG experiments, each user was instructed to spell a sequence of Japanese \emph{kana} characters presented in random order as in the classical P300 response based oddball paradigm~\cite{bciBOOKwolpaw}. The following two types of experimental sessions were performed for each user in the EEG experiments: training session - using only the five Japanese vowels (\emph{a, i, u, e,} and \emph{o}) in a simple single--step--spelling set--up in order to train a stepwise linear discriminant analysis (SWLDA) classifier~\cite{krusienski2006} to discriminate the $P300$ responses (a binary classification mode of \emph{target} versus \emph{non--target} brainwave evoked responses); a copy spelling test session - to test online the two--step--input Japanese \emph{kana} characters speller paradigm. As a task in the test experiment four Japanese words (\emph{ho--ta--ru, na--ma--e, fu--ne--o--mo--tu} and \emph{shi--ra--yu--ki--hi--me}) were spelled by the users.
During the online BCI experiments, the EEG signals were captured with eight active electrodes \emph{g.LADYbird} by g.tec medical instruments GmbH, Austria. The electrodes were connected to an EEG amplifier \emph{vAmp} by Brain Products GmbH, Germany. The electrodes were attached to the following head locations \emph{Cz, CPz, P3, P4, Cp5, Cp6 F3} and \emph{F4}, as in the 10/10 extended international system. 
The ground and reference electrodes were attached at \emph{FCz} and the left earlobe head locations respectively. The captured online EEG signals were processed by the in--house enhanced  \emph{BCI2000} application~\cite{yoshihiroANDtomekAPSIPA2013} using SWLDA classifier with features drawn from $0 \sim 800$~ms ERP intervals. The sampling frequency was set to $500$~Hz, a high--pass filter at $0.1$~Hz and the low--pass filter at $40$~Hz, with an electric power line interference rejection notch filter set in a band of $48 \sim 52$~Hz to avoid sub--harmonic interferences. The sound stimuli were presented randomly with an inter--stimulus--interval (ISI) set to $300$~ms. Each sound stimulus had length of $300$~ms either. The brainwave response signals (ERPs) in each classification step were averaged $20$~times (in the classifier training sessions) and only $10$~times (in the testing two--step--input spelling paradigm).
\begin{figure}[!Hb]
	\centering
	\vspace{-0.5cm}
    \includegraphics[width=\linewidth]{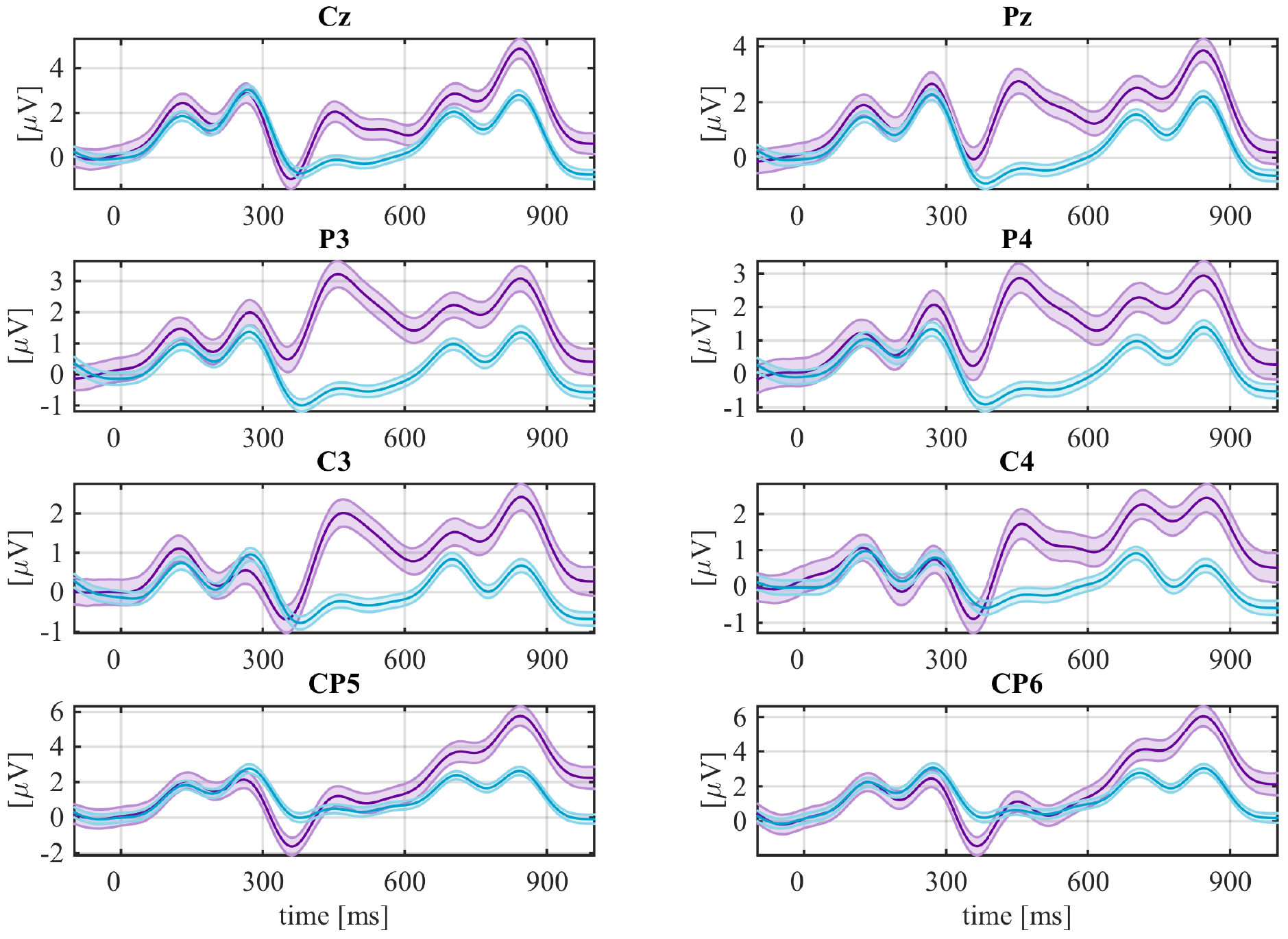}
    \caption{The grand mean averaged ERP brain responses of all users participating in the reported online two--step--input BCI--speller experiments collected within the training sessions. The purple middle (mean) lines depict the ERPs to targets (note the clear P300 positive deflections after about $350$~ms and lasting till about $900$~ms in this study). The blue lines represent the non--targets averaged brainwaves, respectively. Normalized standard errors are depicted with light color intervals surrounding the mean responses.}\label{figures/FullTSH_ALL_P300}
    \vspace{-0.65cm}
\end{figure}
\begin{table}[!Ht]
	\begin{center}
	\caption{Spelling accuracies of the two--step--input Japanese characters paradigm obtained from the online BCI  experiments with a theoretical chance level of $1.7\%$. The averaged scores were $42.76\%, 31.77\%$ and $28.78\%$, for female,  male~$\rightarrow$~female and female~$\rightarrow$~male voice patterns, respectively.}\label{tab:FullTSH_chracters}
	\begin{tabular}{|c|c|c|c|c|c|c|}
	\hline
	User number 	& Voice pattern	& First step accuracy & Final (second step) accuracy \\
	\hline
			&  female	                                 & $53.0\%$	& $53.0\%$ \\ 
	$\#1$ 	& male $\rightarrow$ female	        & $23.5\%$	& $5.9\%$  \\ 
	 		& female $\rightarrow$ male	        & $35.3\%$	& $17.6\%$  \\ 			
	\hline
	        &  female	        & $17.6\%$	& $0.0\%$ \\ 
	$\#2$ 	& male $\rightarrow$ female  	& $11.8\%$	& $5.9\%$  \\ 
	 		& female $\rightarrow$ male	& $0.0\%$	& $0.0\%$  \\ 				
	\hline
			&  female	        & $53.0\%$	& $53.0\%$ \\ 
	$\#3$ 	& male $\rightarrow$ female 	& $76.5\%$	& $47.1\%$  \\ 
	 		& female $\rightarrow$ male 	& $53.0\%$	& $53.0\%$  \\ 				
	\hline
			&  female	        & $0.0\%$	& $33.3\%$ \\ 
	$\#4$ 	& male $\rightarrow$ female	& $23.5\%$	& $17.6\%$  \\ 
	 		& female $\rightarrow$ male	& $11.8\%$	& $5.9\%$  \\ 			
	\hline
			&  female	        & $70.6\%$	& $47.1\%$ \\ 
	$\#5$ 	& male $\rightarrow$ female	& $76.5\%$	& $17.6\%$  \\ 
	 		& female $\rightarrow$ male 	& $82.4\%$	& $64.7\%$  \\ 				
	\hline
			&  female           & $47.1\%$	& $35.3\%$ \\ 
	$\#6$ 	& male $\rightarrow$ female	& $35.3\%$	& $23.5\%$  \\ 
	 		& female $\rightarrow$ male 	& $47.1\%$	& $23.5\%$  \\ 			
	\hline
	        &  female	        & $76.5\%$	& $76.5\%$ \\ 
	$\#7$ 	& male $\rightarrow$ female	& $70.6\%$	& $53.0\%$  \\ 
	 		& female $\rightarrow$ male 	& $35.3\%$	& $23.5\%$  \\ 				
	\hline
			&  female	        & $82.4\%$	& $70.6\%$ \\ 
	$\#8$ 	& male $\rightarrow$ female	& $94.1\%$	& $82.4\%$  \\ 
	 		& female $\rightarrow$ male	& $82.4\%$	& $64.7\%$  \\ 				
	\hline
			&  female	        & $29.4\%$	& $23.5\%$ \\ 
	$\#9$ 	& male $\rightarrow$ female	& $47.1\%$	& $23.5\%$  \\ 
	 		& female $\rightarrow$ male	& $35.3\%$	& $23.1\%$  \\ 			
	\hline
			&  female	        & $47.1\%$	& $35.3\%$ \\ 
	$\#10$ 	& male $\rightarrow$ female	        & $53.0\%$	& $41.2\%$  \\ 
	 		&female $\rightarrow$ male	& $41.2\%$	& $11.8\%$  \\ 				
	\hline
\end{tabular}
\end{center}
\vspace{-1.0cm}
\end{table}

\section{Results}

The psychophysical experiments conducted with all the ten users confirmed no significant differences among the behavioral reaction times (button presses) to the used spatial stimuli.
The EEG experiments conducted with the ten participating users with the proposed Japanese \emph{kana} characters in two--step--input speller using spatial auditory BCI paradigm have been reported below.
The obtained grand mean averaged ERP responses from the ten users has been depicted in Figure~\ref{figures/FullTSH_ALL_P300}. 
The P300 responses have been clearly depicted for targets (purple lines) in the ranges of $350 \sim 600$~ms. The grand mean averaged result confirmed the validity of the proposed saBCI--speller paradigm settings and configuration. The clear P300 responses in comparison to the non--targets (blue lines in Figure~\ref{figures/FullTSH_ALL_P300}) validated the possibility to discriminate those ERPs using machine learning approaches and especially the proposed application of the SWLDA. 
The Japanese \emph{kana} characters set in two--step--input spelling accuracy results of the ten participating users are summarized in Table~\ref{tab:FullTSH_chracters}. The scores of the users' majority were above the two--step--input BCI--speller theoretical chance level of $1.7\%$. 
The female, male~$\rightarrow$~female and female~$\rightarrow$~male stimulus modalities resulted with non--significant differences of means as tested with the pairwise \emph{t-tests.} The ten tested saBCI--speller users reported preferences for the mixed male and female voice patterns due to more easy orientation in the two--step interface. 
The results of the conducted online EEG experiments have shown the feasibility of the Japanese \emph{kana} characters set in two--step--input saBCI--speller paradigm. Also we confirmed that the accuracy of each modality had no significant differences. A video documenting  the online saBCI--speller experiment is available at~\cite{youtubeSABCIfebruary2015}.

\section{Conclusions}

In this paper, we reported the results obtained with a two step input Japanese syllabary saBCI--speller paradigm using $45$ Japanese \emph{kana} characters allowing for a smooth syllabary--based communication.

The results obtained allowed the ten participating users to spell short words using two steps, as reported in the Table~\ref{tab:FullTSH_chracters}. The mean results of all the users were above the theoretical chance level of the proposed two--step--input saBCI--speller. However, zero accuracies were also observed at sporadic cases.
The preliminary yet encouraging results presented call for more research into two--step input spatial auditory BCI--speller paradigms. The next research steps will include the spatial auditory stimulus optimization for handicapped or bedridden subjects, who cannot utilize a full surround sound acoustic environment. We plan to continue this line of research in order to apply the method in the online BCI application for patients suffering from locked--in syndrome.


\end{document}